\documentclass[%
 reprint,
 amsmath,amssymb,
 aps,
]{revtex4-2}

\usepackage{comment}
\usepackage[dvipsnames]{xcolor}
\usepackage{graphicx}% Include figure files
\usepackage{dcolumn}% Align table columns on decimal point
\usepackage{bm}% bold math
\usepackage{hyperref}% add hypertext capabilities
\usepackage{amsmath,amsfonts,amssymb}
\usepackage{float}
\begin{document}

\preprint{APS/123-QED}

\title{Revisiting The Gravitational Mirroring In Presence of Compact Objects}% Force line breaks with \\

\author{Bikramarka S Choudhury}
\email[Email: ]{ bikramarka@gmail.com}
\affiliation{Department of Mathematics, Jadavpur University, Kolkata 700032, West Bengal, India}

\author{Aritra Sanyal}
\email[Email: ]{aritrasanyal1@gmail.com}
\affiliation{Department of Mathematics, Jadavpur University, Kolkata 700032, West Bengal, India}

\author{Md Khalid Hossain}
\email[Email: ]{mdkhalidhossain600@gmail.com}
\affiliation{Department of Mathematics, Jadavpur University, Kolkata 700032, West Bengal, India}

\author{Farook Rahaman}
\email[Email: ]{rahaman@associates.iucaa.in}
\affiliation{Department of Mathematics, Jadavpur University, Kolkata 700032, West Bengal, India}

\begin{abstract}
We propose a novel concept of astrophysical mirroring in the schwarzschild framework, which emerges as a direct consequence of gravitational lensing effects occurring in the immediate vicinity of extremely dense massive objects within spacetime. Through rigorous theoretical calculations and numerical ray-tracing analysis, we demonstrate that sufficiently compact astrophysical objects possess the capability to induce such extreme curvature in spacetime that the resulting gravitational field can bend light rays to extraordinary degrees, creating what we term a "reflection image" or mirror-like appearance of the source in distant regions of space. We discuss the theoretical framework as well as the observational consequences of this phenomenon.
\end{abstract}

\maketitle
\section{Introduction}

Einstein's general theory of relativity fundamentally altered our understanding of gravitation by associating the presence of mass and energy with curvature in spacetime \cite{einstein1916}. A direct consequence of this curvature is that freely moving particles and light rays no longer follow Euclidean straight lines, but rather geodesics—curves determined by the spacetime geometry itself \cite{wald1984,carroll2004}. In the flat spacetime of special relativity, straight lines are identified as the shortest distance between two points; by contrast, in curved spacetime, the notion of "straightness" is generalized through geodesics, which represent either shortest or extremal paths \cite{perlick2000,petters2001}. For null trajectories in particular, this principle governs the propagation of light \cite{perlick2004}. Thus, a light ray emitted from a source does not traverse in a constant direction in the Euclidean sense but instead follows a geodesic path determined by the gravitational field of surrounding matter and energy \cite{schneider1992}.

Einstein's field equations admit a variety of exact solutions representing different astrophysical scenarios \cite{stephani2003}. Of primary importance is the Schwarzschild solution \cite{schwarzschild1916}, the static spherically symmetric vacuum solution, which serves as the foundation for understanding non-rotating compact objects. If the radius of a massive body contracts within its associated Schwarzschild radius, spacetime exhibits a black hole structure; the corresponding surface (the event horizon) and singularity structure have been explored extensively \cite{penrose1965,hawking1973}. This extreme trapping of geodesics illustrates the maximal effect of gravitational curvature. More general solutions include the Kerr metric, which models rotating black holes and introduces the phenomenon of frame dragging \cite{kerr1963,bardeen1972}, and hypothetical wormhole solutions, which mathematically represent nontrivial topologies connecting distant regions of spacetime \cite{morristhorne1988,visser1995,lobo2017}. While each solution reveals unique features, the study of compact objects remains particularly compelling, as their immense densities lead to spacetime geometries with the most pronounced influence on photon motion and on observable signatures such as shadows and photon rings \cite{falcke2000,broderick2009,akiyama2019}.

The degree of bending depends directly on the spacetime curvature, which in turn is stronger near denser objects \cite{eddington1920}. Among the most prominent manifestations of this curvature is gravitational lensing, wherein the path of light is deflected by a massive body lying in the line of sight between a source and an observer \cite{refsdal1964,walsh1979}. Depending on the strength of curvature, one distinguishes between different regimes including weak lensing \cite{bartelmann2001,kilbinger2015,r1,X1,x22}, strong lensing with looping geodesics \cite{bozza2002,virbhadra2000,oguri2019,r2}, and microlensing \cite{paczynski1986,wambsganss2006,gaudi2012}, all of which play crucial roles in astrophysical and cosmological observations. While conventional studies frequently employ approximate lensing equations to model deflection angles and magnifications \cite{bozza2003,eiroa2002}, in this work we center our analysis directly on the null geodesics of curved spacetimes in order to probe alternative phenomena connected to lensing \cite{perlick2015,virbhadra2008}.

In this context, we focus on the implications of null geodesics near compact objects for the bending and possible return of light rays \cite{r3,r4,r5,r6}. Our proposal is that in certain configurations, photon trajectories may be curved by such an amount that the light returns toward its original source. This mechanism leads to the astrophysical mirroring effect, wherein the compact body acts effectively like a mirror situated in spacetime, generating replica images of a source at distinct observational positions \cite{sereno2002}. Unlike conventional gravitational lensing, where the source, lens, and observer occupy a specific alignment producing multiple images, the mirroring phenomenon derives from null geodesic paths that spatially retrace to the source itself \cite{r7,r8,r9,r10,r11}. In our analysis, we restrict attention to Schwarzschild spacetime through geodesic study, motivated by its applicability to dense stars or non-rotating compact remnants, and explicitly demonstrate the existence of geodesics realizing this effect.

Recent studies of horizon-scale imaging emphasize the transition from qualitative ring-like features toward quantitative photon-ring characterization. Detailed ray-tracing analyses show that the first photon-ring peak may emerge as a relatively geometry-sensitive observable at higher angular resolution and observing frequencies, though current measurements remain limited by emission-model degeneracies \cite{Urso2025Equatorial}. In parallel, hierarchical Bayesian imaging frameworks have been developed to directly quantify uncertainties in ring width and brightness from VLBI data, providing a systematic route toward photon-ring inference with future EHT and ngEHT observations \cite{Tiede2025Bayesian}. Additional work explores how microlensing effects and extended gravity scenarios could imprint measurable distortions on photon-ring morphology, highlighting the potential of next-generation interferometry for precision tests of strong-field gravity \cite{Verma2025Microlensing,Yue2025ExtendedGravity}.

The general idea of spacetime mirroring was previously discussed by W.~M.~Stuckey~\cite{stuckey1993}; however, the scope and methodology of the present work are substantially different. In Ref.~\cite{stuckey1993}, the phenomenon—referred to as \emph{boomerang emission}—is formulated as a geometric consequence of the Schwarzschild spacetime, with a detailed analysis of photon emission and reception angles. That study predicts the appearance of rings of returning photons and computes their reception angles for an observer moving from spatial infinity toward the photon sphere.

In contrast, we adopt a physically motivated and observationally relevant perspective. We use the physical arguments for an observer located at Earth or in its vicinity, and the mirroring phenomenon is investigated through numerical ray tracing of null geodesics in the Schwarzschild spacetime. We suggest that such effects can arise in the vicinity of highly compact objects in general. The emphasis is placed on the physical mechanism, the multiplicity of images, and possible observational consequences, particularly in view of recent developments in photon rings and horizon-scale imaging~\cite{gralla2019,gralla2020,johnson2019,akiyama2021}. We suggest that, although highly theoretical, the present framework opens a new and timely avenue for observational astronomy.

The structure of the paper is as follows. In Sec. \ref{2}, we derive the geodesic equations for null trajectories in schwarzschild spacetime and establish the mathematical framework for photon propagation. In Sec. \ref{3}, we present numerical integrations of these geodesics along with ray-tracing analyses that illustrate the possible paths of photons under the influence of strong curvature. Sec. \ref{4} introduces the notion of astrophysical mirroring in detail, expanding upon the mechanism by which light rays may return to their source. The implications of image multiplicity arising from successive retracings are discussed in Sec. \ref{5}. In Sec. \ref{6}, we compare and contrast this effect with closed timelike curves to clarify the distinction between the two phenomena. Potential observational tests and signals relevant to astrophysical mirroring are considered in Sec. \ref{7}. Finally, in Sec. \ref{8}, we summarize the key results of our study and discuss the broader implications for gravitational physics and observational astrophysics.

\section{Geodesic Equation for Null Geodesics in Schwarzschild Spacetime} \label{2}

The trajectory of light rays in schwarzschild spacetime is determined by the null geodesic equation. The schwarzschild metric in natural units $ (G = c = 1) $ is
\begin{equation}
ds^2 = -\left(1 - \frac{2M}{r}\right) dt^2 + \frac{dr^2}{\left(1 - \frac{2M}{r}\right)}  + r^2 \left(d\theta^2 + \sin^2 \theta\, d\phi^2 \right) .
\end{equation}

For the analysis of light paths (null geodesics), we set $$ ds^2 = 0 $$ and restrict to the equatorial plane ($ \theta = \frac{\pi}{2} $), so $ d\theta = 0 $.

The geodesic equations admit two conserved quantities due to spacetime symmetries:
\begin{align}
E &= \left(1 - \frac{2M}{r}\right)\frac{dt}{d\lambda}, \\
L &= r^2 \frac{d\phi}{d\lambda},
\end{align}
where $E$ and $L$ are the energy and angular momentum per unit mass (for null geodesics), and $  \lambda  $ is an affine parameter along the light path.

Substituting these into the metric and using $$ ds^2 = 0 $$ gives
\begin{equation}
0 = -\left(1 - \frac{2M}{r}\right)\left(\frac{dt}{d\lambda}\right)^2 + \left(1 - \frac{2M}{r}\right)^{-1}\left(\frac{dr}{d\lambda}\right)^2 + r^2 \left(\frac{d\phi}{d\lambda}\right)^2.
\end{equation}

Using the expressions for $  E  $ and $  L  $, we obtain:
\begin{equation}
\left(\frac{dr}{d\lambda}\right)^2 = E^2 - \left(1 - \frac{2M}{r}\right) \frac{L^2}{r^2}.
\end{equation}

We introduce the variable $  u = 1/r  $, so that
$$
\frac{dr}{d\lambda} = \frac{dr}{d\phi} \frac{d\phi}{d\lambda} = -\frac{1}{u^2} \frac{du}{d\phi} \frac{d\phi}{d\lambda} = -\frac{L}{u^2} \frac{du}{d\phi}.
$$
Squaring and substituting into the geodesic equation gives
$$
\left(\frac{dr}{d\lambda}\right)^2 = L^2 \left(\frac{du}{d\phi}\right)^2.
$$
Thus the equation becomes
$$
L^2 \left(\frac{du}{d\phi}\right)^2 = E^2 - \left(1 - 2 M u \right) L^2 u^2.
$$
Dividing by $  L^2  $ and reordering terms, we find
\begin{equation}
\left(\frac{du}{d\phi}\right)^2 + u^2 - 2Mu^3 = \left(\frac{E}{L}\right)^2.
\end{equation}

For a null geodesic, the standard form of the orbit equation is obtained by differentiating $  u  $ with respect to $  \phi  $:
\begin{equation} \label{geodesic}
\frac{d^2 u}{d\phi^2} + u = 3 M u^2.
\end{equation}

This differential equation entirely governs the trajectory of light rays in the schwarzschild spacetime. The nonlinear term $  3 M u^2 $  encodes the general relativistic correction responsible for phenomena such as the gravitational bending of light near massive bodies.

\section{Numerically Integrating and Ray-tracing} \label{3}

\begin{figure}[H]
    \centering
    \includegraphics[width=\linewidth]{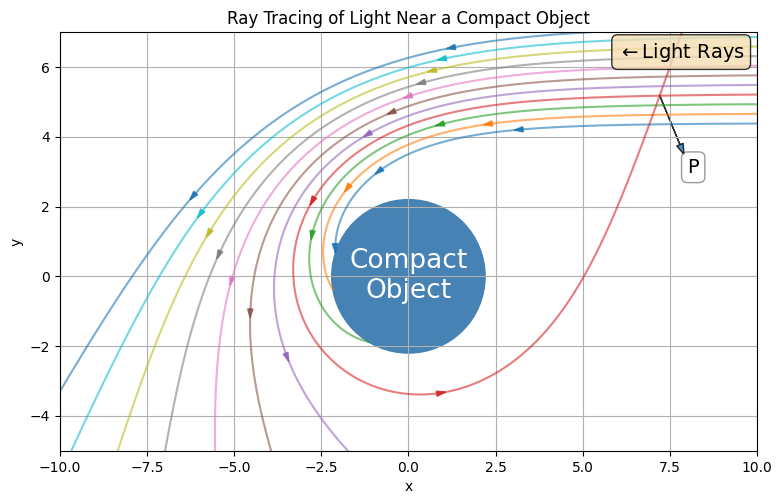}
    \caption{Light tracing back to the point of emission. Suppose the point of emission is the self-intersecting point 'P' of the 'red line.' Then after traveling around the compact object, the light path spatially traces back to the source point, thereby creating a mirror image of the point at a distant region of space. We have taken the mass of the compact object to be $1\,M_\odot$ and radius of the object 2.2$\,km$ which is greater than the schwarzschild radius. In the above plot, the units of $x$ and $y$ axes are in $km$.}
    \label{fig:mirror}
\end{figure}

\begin{figure}[h]
    \centering
    \includegraphics[width=\linewidth]{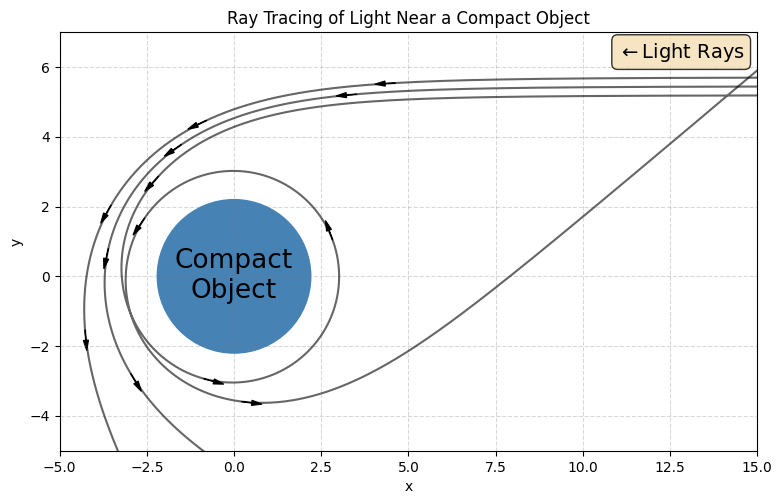}
    \caption{Secondary image formation due to the light looping around it twice. If we suppose the point of emission of light is the self-intersecting point of the line that loops around the compact object, then after looping twice, the light traces back to the source. Hence, this certain path of light creates a secondary image of the source at a distant region of space. We have taken the mass of the compact object to be $1\,M_\odot$ and radius of the object 2.2$\,km$ which is greater than the schwarzschild radius. In the above plot, the units of $x$ and $y$ axes are in $km$.}
    \label{fig:secondary}
\end{figure}

The equation \eqref{geodesic} is a non-linear second order differential equation which is difficult to solve analytically. There are methods like the perturbation method or weak field approximation, etc., to get to the solutions to this differential equation. However, in our study we consider numerical evaluation of the integral through the Runge-Kutta 4th order method ($RK4$ method) and plot the null geodesic for certain paths in a cartesian coordinate plane. We use the ray-tracing method in Python for achieving our results given in Fig. \ref{fig:mirror} and Fig. \ref{fig:secondary}.

\section{The concept of Mirroring}\label{4}

One of the central principles of general relativity is that mass and energy cause spacetime to curve. This curvature governs how objects---and even light---move within the universe. When a massive object, such as a star or black hole, exists, it warps the fabric of spacetime; this phenomenon is fundamental to the Einstein field equations.

Now the denser the object, the more light is bent while passing near the object. The spacetime outside a static spherically symmetric mass distribution is given by the schwarzschild metric. Now from the schwarzschild metric, we see if any object is so dense that its radius is below the schwarzschild radius, then it creates an event horizon at the schwarzschild radius which is the defining characteristic of schwarzschild black hole. In geometrical units, if the mass of an object is given as $M$ units, then the schwarzschild radius is $R_s=2M$ units. There is another radius which is very interesting to note. It's called the photon sphere, and the radius is $R_p=3M$ units.

At this distance from the center, gravity is strong enough that light itself can orbit the object along unstable, circular paths. These photon orbits are not stable---any small disturbance will cause the photon to escape or fall inward---but the photon sphere is crucial to understanding the ``shadow'' or silhouette observed in real black hole images.

It is important to highlight that the existence of a photon sphere does not require an event horizon. If the object's radius is between the Schwarzschild radius ($2M$) and the photon sphere ($3M$), it is not a black hole since it lacks an event horizon, but a photon sphere can still exist. This means any sufficiently dense astrophysical object can support unstable circular orbits for light near its surface. Thus, if light moves very close to such an object, its path can be so strongly bent that it loops around in a closed circle.

As we examine how light bends near these dense objects, a pattern emerges: far from the mass, the deflection is small; closer in, the bending intensifies. At the photon sphere, the deflection is so extreme that photons can, in principle, orbit indefinitely (though such orbits are unstable). This enables fascinating possibilities: if a light source is positioned just right, photons emitted from it can return to their origin after encircling the compact mass. For an observer at that location, this would appear as a ``mirror image'' from the past, since returning photons reveal earlier states of the emitter.

It is interesting to note that any object that is sufficiently dense can cause nearby light to loop back or undergo significant lensing. For example, if a supermassive compact object, like the core of a galaxy, lies between us and another luminous source (such as a distant galaxy), its gravity can bend the light to create multiple images, arcs, or, in extreme cases, mirror-like effects. This mechanism underlies gravitational lensing, which lets us observe distant sources magnified, distorted, or even multiplied by massive foreground objects.

\section{Image Multiplicity}\label{5}

The image formation process in the scenario under consideration does not correspond to a single, isolated image of the source. Instead, due to the strong gravitational field of the compact object, multiple images are generated. The mechanism is as follows.  

The first image is formed by the primary bending of light: photons emitted from the source are deflected by the gravitational field and return directly to the observer after a single pass around the object. This is the most prominent and easily identifiable image.  

The second image arises when the photon trajectory involves an additional detour: instead of simply bending once, the photon executes a loop around the compact object before eventually reaching the observer. This additional winding increases the path length and hence the time delay relative to the first image. The figure below represents the scenario where light emitted from the source has returned to the source point after winding once around the compact object, therefore creating the second image.

Similarly, the third image corresponds to photons that loop twice around the compact object before escaping towards the observer. Each successive image is produced by photons undergoing an increasing number of windings around the gravitational lens.  

In principle, this sequence continues indefinitely. That is, there exists a theoretically \emph{countably infinite set of images}, each associated with a photon path that loops around the compact object $n$ times (for $n = 0, 1, 2, \dots$). The angular positions of these higher-order images converge rapidly towards the photon sphere of the compact object. As a result, they appear in very close proximity to one another, almost overlapping in the observer’s sky.  

From an observational standpoint, this clustering implies that the higher-order images are extremely difficult to resolve individually. Their separation is far below current instrumental resolution, and their flux also decreases sharply with increasing winding number due to both geometrical path elongation and redshift effects. Consequently, while the theoretical prediction is that of an infinite image sequence, in practice only the first one or two images might be detectable, with the remainder forming an unresolved accumulation near the photon sphere.

This mirroring effect does not produce a single, isolated image but rather generates a complex pattern of multiple images, each corresponding to light rays that have undergone different numbers of orbital loops around the lensing object before escaping toward the observer. The primary image results from photons that experience a single deflection event, while secondary and higher-order images arise from light that completes one, two, or more partial orbits around the compact mass before reaching the detector. In principle, this sequence of images continues indefinitely, creating a theoretically infinite series of progressively fainter reflections, though observational limitations restrict the practical detection to only the brightest few images in the sequence.

\section{The Definition And Clarification}
\label{sec:mirroring_definition}

In this work, \emph{astrophysical mirroring} is defined as a special case scenario of strong-field gravitational lensing effect arising from the behavior of null geodesics in the vicinity of a sufficiently compact object. The phenomenon refers specifically to photon trajectories that are emitted from a source, undergo extreme gravitational deflection due to spacetime curvature, and return to the spatial location of the source after executing one or more partial or complete windings around the compact object.

We emphasize that astrophysical mirroring does \emph{not} correspond to a physical reflection process or to an optical image inversion analogous to reflection from a material mirror. Instead, it is a purely geometric effect governed by the properties of null geodesics in curved spacetime. The apparent ``mirror-like'' behavior arises from the existence of near-critical photon trajectories that experience large deflection angles in the strong-field regime. From an observational perspective, astrophysical mirroring therefore manifests not as a literal inverted image, but as a sequence of higher-order lensed images whose angular positions accumulate near the photon sphere.

\begin{comment}

\textcolor{purple}{
The propagation of photons in Schwarzschild spacetime is characterized by the impact parameter
\begin{equation}
b = \frac{L}{E},
\end{equation}
where \(E\) and \(L\) are the conserved energy and angular momentum of the photon, respectively. For impact parameters approaching the critical value associated with the photon sphere,
\begin{equation}
b_c = 3\sqrt{3}M,
\end{equation}
the total deflection angle increases rapidly and diverges logarithmically as \(b \to b_c\). As a result, small variations in the emission angle at the source lead to large changes in the deflection angle and, consequently, in the angular position of the image on the observer’s sky.
}

\textcolor{purple}{
This highly non-linear mapping between emission angle and observed image position enables a subset of photon trajectories to return to the spatial vicinity of the source, thereby generating additional gravitationally lensed images associated with successive windings around the compact object. From an observational perspective, astrophysical mirroring therefore manifests not as a literal inverted image, but as a sequence of higher-order lensed images whose angular positions accumulate near the photon sphere.
}

\end{comment}

Astrophysical mirroring is thus best understood as a specific realization of strong-field gravitational lensing in which null geodesics retrace to the source location due to extreme spacetime curvature, rather than as an optical reflection phenomenon.

\section{Strong Lensing Effect With Deflection Angle $(2k+1)\pi$}

In the schwarzschild metric, restricting photon motion to the equatorial plane $(\theta=\pi/2)$ and using the conserved energy $E$ and angular momentum $L$, the null geodesic equation can be written as
\begin{equation}
\left(\frac{dr}{d\phi}\right)^2
=
r^4\left[
\frac{1}{b^2}
-
\frac{1}{r^2}\left(1-\frac{2M}{r}\right)
\right],
\end{equation}
where $b=L/E$ is the impact parameter.

The total deflection angle of light is then
\begin{equation}
\alpha(b)
=
2\int_{r_0}^{\infty}
\frac{dr}
{r^2\sqrt{\frac{1}{b^2}-\frac{1}{r^2}\left(1-\frac{2M}{r}\right)}}
-\pi,
\end{equation}
with $r_0$ denoting the distance of closest approach.

In Schwarzschild spacetime, photons admit an unstable circular orbit at the photon sphere,
\begin{equation}
r_{\mathrm{ph}} = 3M,
\end{equation}
corresponding to the critical impact parameter
\begin{equation}
b_c = 3\sqrt{3}\,M.
\end{equation}

As $b \to b_c^+$, the deflection angle diverges logarithmically, allowing photons to wind around the compact object multiple times. In the strong deflection limit, the bending angle can be approximated by \cite{strong_lensing}:
\begin{equation}
\alpha(b)
\simeq
-\ln\!\left(\frac{b}{b_c}-1\right)
+
\ln\!\left[216(7-4\sqrt{3})\right]
-\pi.
\end{equation}

Thus, the lensing effect where the photon retraces a path straight back corresponds to a total deflection angle
\begin{equation}
\alpha = (2k+1) \pi, \quad k=0,1,2,3,...
\end{equation}

Solving $\alpha(b)=\pi$ in the strong deflection regime yields
\begin{equation}
\frac{b_{\pi}}{b_c}-1
\simeq
\exp\!\left[
-\left(2(k+1)\pi - \ln\!\left[216(7-4\sqrt{3})\right]\right)
\right],
\end{equation}
which implies
\begin{equation}
b_{\pi} \approx 1.0006\, b_c, \quad \text{for $k=1$}.
\end{equation}

The case $k=1$ corresponds to the \textit{first image} as defined in the previous sections of our work. 
This configuration provides an estimate of the impact parameter in the strong--lensing regime, 
where the null geodesic is bent in such a manner that it effectively retraces its path back to the observer. 
The choice of deflection angles given by odd multiples of $\pi$ is motivated by the presence of an extremely compact 
and highly dense object with a smaller radii located at a sufficiently large distance. 
In this limit, the bending of light closely mimics a direct redirection toward the observer, 
analogous to reflection from a mirror. 
This mirror--like behavior, arising purely from the extreme curvature of spacetime and the resulting bent null trajectories, 
constitutes the central physical insight of our work.

\section{Difference with closed time like curve}\label{6}

We would like to emphasize one important clarification regarding the interpretation of our results. At first sight, the trajectories we obtain might appear to suggest the existence of closed timelike geodesic curves. However, this is not the case in our present analysis. The framework of our study is restricted to the spatial sector of the geometry. Consequently, the closed structures that we depict and analyze are spatial loops, not closed geodesics in spacetime.

If the temporal dimension is properly incorporated into the description, these apparent loops no longer remain closed. Instead, they extend along the time axis, thereby forming spiral-like structures in spacetime. In other words, the “closure” we observe in the spatial projection is only a visual artifact of suppressing the time coordinate; the full spacetime trajectories remain open when viewed in four dimensions.

Thus, our study should not be interpreted as evidence for the existence of closed timelike curves. Rather, it highlights the spatial manifestation of geodesic motion, which, when embedded in the full spacetime manifold, acquires a spiral character due to its natural extension along the time dimension.

\section{Possible Observational Consequences and Feasibility} \label{7}

\subsection{Brightness of Galactic Center}

An important implication of our framework is its direct relevance to the observed luminosity of galactic centers. It is well established that these regions are characterized by extreme densities, often associated with compact objects or dense stellar clusters. Under such conditions, the mechanism we have described becomes naturally applicable.

In particular, a sufficiently dense galactic core can act as a powerful gravitational lens, bending and redirecting radiation in such a way that a portion of the light is traced back toward the observer. This effect applies both to radiation originating within the galaxy itself and to photons emitted by external astrophysical sources (including our own galaxy) that are deflected by the dense core and redirected into our line of sight. We understand that explaining this from observing the luminosity of the galaxies can be problematic due to the existence of a huge amount of exotic matter in the galaxies. However, our work suggests that in the presence of highly compact objects, the luminosity due to astrophysical bodies present in the galaxy will be even lesser than that observed by us.

Consequently, the extraordinary brightness of galactic centers cannot be attributed solely to intrinsic emission processes such as accretion dynamics, stellar radiation, or non-thermal activity. A significant contributory factor arises from the cumulative effect of redirected radiation: part of the light from the galaxy’s own stellar and interstellar sources, together with light from background objects, is effectively focused back to the observer.

Thus, the luminous appearance of galactic centers should be understood as the outcome of both intrinsic energetic processes and the strong gravitational redirection of light by their dense cores. Our study therefore provides a natural explanatory framework in which gravitational light bending plays a central role in shaping the observed radiative properties of  galaxies.

\subsection{Angular Resolution and EHT Constraints}

The characteristic angular scale associated with photon-sphere-related features is of order
\begin{equation}
\theta_{\rm ph} \sim \frac{3\sqrt{3}M}{D},
\end{equation}
where $D$ is the distance to the compact object. For Sagittarius A*, this corresponds to an angular scale of approximately $50~\mu\mathrm{as}$, comparable to the resolution of the Event Horizon Telescope (EHT). However, the angular separation between mirrored images and standard photon-ring substructures is significantly smaller than current EHT resolution limits. As a result, individual mirroring images cannot be resolved with present instruments and would, at most, contribute subtly to the brightness distribution near the photon ring.

\subsection{Isolated Black Holes}

This phenomenon implies that, if an isolated black hole exists, some amount of light should necessarily be present in its immediate vicinity due to gravitational mirroring, even when the spacetime behind the black hole is completely dark. 

The extreme curvature of spacetime around the black hole effectively acts as a mirror, bending and redirecting background starlight or ambient radiation back toward the observer, thereby producing a faint but nonzero luminous signature. 

This offers an important observational handle: several isolated dark, compact objects identified through microlensing by the \textit{Hubble Space Telescope} may be more naturally interpreted as neutron stars or other stellar remnants, rather than entirely "invisible" black holes, since genuine black holes would be expected to leave traces of mirrored light in their surroundings.

However, this effect is likely to be extremely weak and challenging to resolve with current observational capabilities, and its detection may therefore require next-generation very-long-baseline interferometry or indirect statistical techniques rather than direct imaging.

Consequently, astrophysical mirroring not only helps disfavor black holes as explanations for many microlensed \textit{Hubble} objects but also provides guidance on the signatures to search for when attempting to identify isolated black holes drifting through interstellar space.

\subsection{Redshift}
To elucidate the redshift behavior, let us consider a scenario in which light is emitted by an observer situated within a gravitationally bound cluster and is subsequently mirrored back by a distant compact object. As the photon initially propagates outward from the gravitational potential well of the cluster, it undergoes a gravitational redshift. During its inward journey toward the compact object, the photon is gravitationally blueshifted due to the increasing curvature of spacetime. After the mirroring effect, the photon again experiences a gravitational redshift while moving away from the compact object, followed by a final gravitational blueshift as it falls back into the gravitational potential of the observer’s cluster.

Consequently, the gravitational redshift and blueshift contributions acquired along the photon’s trajectory cancel exactly, resulting in no net gravitational frequency shift. In contrast, the cosmological redshift accumulates continuously from the time of emission and remains unaffected by the photon’s reversal of direction. Since the photon effectively traverses the distance between the observer and the compact object twice, the observed cosmological redshift corresponds to the expansion of the universe over this doubled path length.

\section{Conclusion and Discussion}\label{8}

In summary, the intense curvature of spacetime in the vicinity of compact objects such as black holes produces a striking phenomenon known as gravitational mirroring. Light rays passing near these regions are bent so strongly that they can circle the object, giving rise to multiple or distorted images of distant sources. Rather than being a mere optical illusion, this effect provides astrophysicists with a natural laboratory to probe the characteristics of otherwise unseen objects and to examine the validity of general relativity under extreme gravitational conditions. \\

Our numerical ray-tracing analysis confirmed that photons can undergo successive loops, producing a theoretically infinite sequence of images. While practical observational constraints restrict detection to the brightest one or two images, the accumulation of higher-order images near the photon sphere highlights a rich structure in the gravitational lensing regime. Importantly, we clarified that these trajectories, while appearing closed in spatial projections, do not correspond to closed timelike curves but instead form spiral-like structures in full spacetime, preserving causality. From an astrophysical standpoint, our framework offers an alternative perspective on the observed extraordinary brightness of galactic centers. Traditionally attributed to accretion processes, stellar radiation, and non-thermal activity, such luminosity may also be partly explained by gravitational redirection and mirroring of light. Dense galactic cores can focus both internal and external radiation back toward the observer, enhancing the apparent brightness and contributing to the complex structure of galactic emission. This interpretation provides a natural and testable extension of gravitational lensing theory, enriching our understanding of the interplay between compact objects and radiation. \\

\textbf{Remarks:} This work is primarily focused on the schwarzschild metric, which represents one of the simplest forms of spacetime curvature. While this restriction may be regarded as a limitation of our analysis, given the strength of our observational claims, it is important to note that similar mirroring phenomena have also been studied in the rotating Kerr spacetime, as discussed in \cite{cramer1997}. We therefore argue that such effects are not confined to the schwarzschild geometry, but can plausibly extend to more complex curved spacetimes as well. On this basis, we contend that our claims remain valid and are supported by broader theoretical considerations.

\section{Declaration of competing interest}
The authors declare that they have no known competing financial interests or personal relationships that could have appeared to
influence the work reported in this paper.

\section{Acknowledgment}

FR, BSC and AS are thankful to the Inter-University Centre for Astronomy and Astrophysics (IUCAA), Pune, India, for their support. FR, BSC, and AS also gratefully acknowledge academic support from Jadavpur University (JU). BSC acknowledges financial assistance from the UGC. FR thanks ANRF, SERB, Government of India, for their support.

\section{Data Availability}

No datasets were generated or analyzed during the current study.

\nocite{*}

\bibliographystyle{unsrt}
\bibliography{refer}

\end{document}